\documentclass[prd,
a4paper,
nofootinbib,
reprint,
]{revtex4}
\usepackage{graphicx}
\usepackage{amsfonts}
\usepackage{amssymb}
\usepackage{amsmath}
\usepackage{slashed} 
\usepackage{color}
\usepackage{hyperref}
\def\eq#1{{Eq.~(\ref{#1})}}

\def\di{\mbox{d}}

\def\gtap{\ \raisebox{-.4ex}{\rlap{$\sim$}} \raisebox{.4ex}{$>$}\ }

\definecolor{oucrimsonred}{rgb}{0.6, 0.0, 0.0}
\definecolor{persianblue}{rgb}{0.11, 0.22, 0.73}
\definecolor{forestgreen}{rgb}{0.13,0.35,0.13}
 \hypersetup{colorlinks, citecolor=oucrimsonred, linkcolor=persianblue, urlcolor=oucrimsonred}
\def\hhref#1{\href{http://arxiv.org/abs/#1}{#1}} 
\newcommand{\be}{\begin{equation}}
\newcommand{\ee}{\end{equation}}
\newcommand{\bea}{\begin{eqnarray}}
\newcommand{\eea}{\end{eqnarray}}
\newcommand{\nn}{\nonumber}

\newcommand{\U}{\scriptscriptstyle U}
\newcommand{\D}{\scriptscriptstyle D}

\newcommand{\N}{\scriptscriptstyle N}
\newcommand{\uu}{\scriptscriptstyle U\!U}
\newcommand{\dd}{\scriptscriptstyle D\!D}
\newcommand{\UR}{\scriptscriptstyle U_{\!R}}
\newcommand{\DR}{\scriptscriptstyle D_{\!R}}
\newcommand{\UL}{\scriptscriptstyle U_{\!L}}
\newcommand{\DL}{\scriptscriptstyle D_{\!L}}
\newcommand{\R}{\scriptscriptstyle R}
\newcommand{\LL}{\scriptscriptstyle L}

\begin{document}
\title[]{Neutral hadrons disappearing into the darkness
}
\date{\today}
\author{D.\ Barducci$^{\dag\ddag}$}
\author{M.\ Fabbrichesi$^{\ddag}$}
\author{E.\ Gabrielli$^{\ast \ddag \natural \circ}$}
\affiliation{$\dag$Scuola Internazionale di Studi Superiori, via Bonomea 256, 34136 Trieste, Italy}
\affiliation{$^{\ddag}$INFN, Sezione di Trieste, Via  Valerio 2, 34127 Trieste, Italy }
\affiliation{$^{\ast}$Physics Department, University of Trieste, Strada Costiera 11, 34151 Trieste, Italy}
\affiliation{$^{\natural}$NICPB, R\"avala 10, Tallinn 10143, Estonia }
\affiliation{$^{\circ}$Theoretical Physics Department,  CERN, CH-1211 Geneva 23, Switzerland}
\begin{abstract}
\noindent  We study the invisible decay of neutral hadrons in a representative model of the dark sector.
The   mesons 
  $K_L$ and  $B^0$  decay into the dark sector with branching rates that can be at the current experimental limits. The neutron decays with a rate that
could  either explain  the neutron lifetime puzzle (although only for an extreme choice of the  parameters and a fine tuned value of the  masses) or be  just above   the current limit of its invisible decay ($\tau_N^{\tiny \mbox{inv}}  \gtap 10^{29}$ years) if kinematically allowed.
These invisible decays of ordinary matter  provide a novel and promising window into new physics that should be vigorously pursued. 
\end{abstract}
\maketitle
 
 \section{Motivations}
 
The possible existence of a \textit{dark sector}  comprising  particles  that do not couple directly to the Standard Model (SM) states  has been extensively discussed in the literature (see references in~\cite{Raggi:2015yfk} for recent reviews). This dark sector can include many states and these states can interact among themselves by means of new forces. Dark matter, in this framework, is made of all the stable members of  the dark sector with a non-negligible relic density.

If the  dark sector contains sufficiently light states, ordinary matter can and will decay into it  
without leaving any trace. These invisible decay channels are striking and may well be the most conspicuous clue to the existence of the dark sector itself. 

Because of charge conservation, only neutral hadrons can  altogether  decay into the dark sector. The invisible decays of  Kaons and $B$-mesons are of particular interest because  their long lifetimes  provide  appreciable branching rates (BR) even for decays as rare as those into the dark sector.  In addition to these, the case of the neutron stands out both because of the very strong bound on  its invisible decay  and because
of the experimental discrepancy between the lifetime  measured from stored neutrons and that from in-beam decays (for a review, see \cite{Paul:2009md}) which could be explained, as pointed out in~\cite{Fornal:2018eol}, by an invisible decay.  

\section{A model for the dark sector}

We restrict ourselves to a model in which the interaction with ordinary matter is provided by (heavy)   \textit{messenger} states.
This model is taken to be the archetype for a dark sector that can leave a characteristic signature of its interaction with ordinary mater in, among other processes, the invisible decays of hadrons.

The dark sector is made to resemble QED---that is, a  theory of charged fermions. It contains  fermions $Q^{\U_i}$ and $Q^{\D_i}$, where  the index $i$ runs over generations like in the SM, and these dark fermions are charged only under a gauge group $U(1)_{\D}$---a proxy  for more general interactions---with different charges for the  $Q^{\U}$ and $Q^{\D}$ type. The  \textit{dark} photon is massless and directly only couples to the dark sector (in contrast with the case of massive dark photons).
We denote throughout with $\alpha_{\D}=g_{\D}^2/4 \pi$ the $U(1)_{\D}$ fine structure constant.
 
The dark fermions couple to  the SM fermions by means of  Yukawa-like interactions. The Lagrangian contains terms coupling quarks of different generations  with the dark fermions. In general the interaction is not diagonal and, for the quark case, is given by 
\be
{\cal L}  \supset \ g_R \Big\{ 
S^{\U_i \dag }_R \left[ \bar{Q}^{\U_i}_L (\rho^{\U}_R)_{ij} q^j_R \right]  +
S^{\D_i \dag }_R \left[ \bar{Q}^{\D_i}_L (\rho^{\D}_R)_{ij} q^j_R\right]   \Big \} \label{L} \\
 +
 g_L \Big\{
S^{\U_i\dag }_L \left[\bar{Q}^{\U_i}_R (\rho^{\U}_L)_{ij}q^j_L\right]  
+S^{\D_i\dag }_L\left[\bar{Q}^{\D_i}_R (\rho^{\D}_L)_{ij} q^j_L\right]  \Big\}
 + \mbox{H.c.}  
\ee
In \eq{L}, the fields $S^{\U_i,\D_i}_{L}$ and $S^{\U_i,\D_i}_{R}$ are the messenger scalar particles, respectively doublets and singlets of the SM $SU(2)_L$ gauge group as well as  $SU(3)$ color triplets (color indices are implicit in \eq{L}). The various symmetric matrices $(\rho)_{ij}=(\rho)_{ji}$ are the result of  the diagonalization of the mass eigenstates of both the  SM and dark fermions; they provide the generation mixing necessary to have the messengers play a role in flavor physics.  
The messenger  fields are also charged under the $U(1)_{\D}$ gauge interaction, carrying the same charges as the dark fermions they are coupled to. 

 In writing \eq{L} we assume that the SM gauge group $SU(2)_L$ is extended into a  left-right (LR) symmetric $SU(2)_L\times SU(2)_R$ group and follow the approach of \cite{Gabrielli:2016vbb}---to which we refer for further details.  Although we  adopt the LR symmetric model, the low-energy effective theory is not  affected by this choice and is the same as in the model in \cite{Gabrielli:2013jka}. 

The general structure for the gauge invariant Lagrangian   contains a  term involving three scalar messengers and the heavy Higgs $H_R$, a $SU(2)_R$ doublet, coupled as follows (generation index $i$ is implicit this time)
\be
{\cal L}_{3}  \supset \,  \eta_L \tilde{S}^{\U\alpha\dag}_L S^{\D\beta}_L
H^{\dag}_R S^{\D\gamma}_R\varepsilon^{\alpha\beta\gamma}  
 +
\frac{\eta_R}{2} \tilde{S}^{\U\alpha\dag}_R S^{\D\beta}_R H^{\dag}_R S^{\D\gamma}_R\varepsilon^{\alpha\beta\gamma} + \mbox{H.c.} \, ,
\label{interaction}
\ee
provided the $U_{\D}(1)$ dark charges $q^{\U}$ and $q^{\U}$ of, respectively, the messenger
 $S^{\U\dag}_{L,R}$ and  $S^{\D\dag}_{L,R}$ satisfy the relation $q^{\U}=-2q^{\D}$ (as in the case of up- and down-quark QED charges) for $q^{\U}$ normalized to one.
In  \eq{interaction} above the sum over the Greek $SU(3)$ color indices is understood and $\tilde{S}^{i}_{L,R}=i\sigma_2 S^{i \star}_{L,R}$, where $\sigma_2$ is the Pauli matrix of the corresponding SU(2) group.
After the spontaneous breaking of the $SU(2)_R$ gauge symmetry, the $H_R$ vacuum expectation value 
$v_R$ generates a trilinear term involving three scalar messengers entering 
the vertex. The terms in \eq{interaction} play a role in the decays of baryons.

This model  has been used to discuss processes with the emission of dark photons in Higgs physics~\cite{Biswas:2016jsh}, flavor changing neutral currents~\cite{Gabrielli:2016cut}, kaon~\cite{Fabbrichesi:2017vma} and $Z$ boson~\cite{Fabbrichesi:2017zsc} decays. 

\subsection{Dark matter, relic density and galaxy dynamics}

The messenger fields are heavier than the dark fermions; the latter are stable and provide a multicomponent candidate for dark matter whose  relic density depends on the value of their couplings to the $U(1)_{\D}$ dark photons and SM fermions (into which they annihilate) and masses.

 Not all of the dark fermions contribute to the relic density when, as we do here,  the $U(1)_{\D}$ coupling is taken larger than the one in QED.
If they are relatively light, their dominant annihilation is  into dark photons with a thermally averaged cross section   approximately given by
\be
\langle \sigma v_0 \rangle \simeq \frac{\pi \alpha_{\D}^2}{2 m_Q^2}  \label{thermal-x-section1}
\ee
For a  strength $\alpha_{\D} \simeq 0.1$, all fermions with masses  up to around 1 TeV have a large  cross section and their relic density 
\be
\Omega\, h^2 \approx \frac{2.5 \times 10^{-10} \; \mbox{GeV}^{-2}}{\langle \sigma v_0 \rangle}
\ee
 is only a percent of the critical one; it is
roughly $10^{-4}$ the critical one for dark fermions in the 1 GeV range, even less for lighter states.  These dark fermions are not part of dark matter; they have (mostly) converted into dark photons by the time the universe reaches our age and  can only be produced in high energy events like the decays we discuss. 

Heavier (that is, with masses closer to those of the messengers) dark fermions  can be dark matter. The dominant annihilation for these is  into SM fermions via the exchange of a messenger with a thermally averaged cross section   now approximately given by
\be
\langle \sigma v_0 \rangle \simeq  \left( \frac{g^2_{L,R}}{4 \pi} \right)^2 \frac{\pi }{2 m_S^2} \label{thermal-x-section2}
\ee
instead of \eq{thermal-x-section1}.  The critical relic density can be reproduced if, assuming thermal production,
\be
\left( \frac{g^2_{L,R}}{4 \pi} \right)^2 \left( \frac{10 \, \mbox{TeV}}{m_S} \right)^2 \simeq 0.1 \, . \label{relic}
\ee

Although dark matter is interacting via massless dark photons, limits from the collisionless  dynamics of galaxies are satisfied because the light dark fermions have a negligible density in the galaxy (and do not count)   while for the heavy dark fermions the bound on soft scattering~\cite{Ackerman:mha}, which is the strongest, is given (for $N$  dark fermions of mass $m_Q$, $G_N$ being the Newton constant) by
\be
\frac{G_N^2 m_Q^4 N}{8 \alpha_{\D}^2} \left[ \ln \left( \frac{G_N m_Q^2 N}{2 \alpha_{\D}^2} \right) \right] ^{-1} \gtrsim 50 \, .
\ee
The above bound  can easily be satisfied because it is independent of the parameters entering the relic density. In our case, the above bound means that for $\alpha_{\D} \simeq 0.1$ the heavy dark fermions present in the relic density must have masses   larger than 8 TeV. This limit, together with \eq{relic}, defines the allowed space of the parameters, namely, the couplings $g_{L,R}$ must be large but still in the perturbative regime. 


\subsection{Constraints on messenger masses}

There are no bounds on the masses of the dark fermions because of their very weak interaction with the SM states.

The messenger states have the same quantum numbers and spin of the supersymmetric squarks.
At the LHC they are copiously produced in pairs through QCD interactions and decay at tree level into a quark and a dark fermion. The final state arising from their decay is thus the same as the one obtained from the $\tilde q \to q \chi^0_1$ process.
Therefore limits on the messenger masses can be obtained by reinterpreting supersymmetric searches on first and second generation squarks decaying into a light jet and a massless neutralino~\cite{Aaboud:2017vwy}, assuming that the gluino is decoupled. 
In particular we have used the upper limits on the cross section for various squark masses of~\cite{Aaboud:2017vwy} that the ATLAS collaboration provided on {\tt HEPData}. These limits have been used to 
compute the bounds as a function of the messenger mass  using next-to-leading order QCD cross section for squark pair production from the LHC Higgs Cross Section Working Group~\footnote{Available at the web-page~\url{https://twiki.cern.ch/twiki/bin/view/LHCPhysics/SUSYCrossSections}.}.

We take into account the contributions to the total event yield given only by right-handed (degenerate) messengers associated to the first generation of SM quarks, with the others set to a higher mass and thus with a negligible cross section. This correspond to have only 2 light degrees of freedom, which are analogous to $\tilde u_1$ and $\tilde d_1$ in supersymmetry. With  this assumption we obtain a lower bound on their masses of 940 GeV, limit that increases up to 1.5 TeV by assuming that messengers of both chiralities associated to the first and second generation of SM quarks are degenerate in mass.

Interestingly, there remains an open window for having messengers living at a lower mass scale. This occurs when the messengers couple dominantly to top quarks and have a mass around 200 GeV, such that the final state kinematic presents low missing transverse energy due the compression of the spectrum, thus reducing the effectiveness of supersymmetric searches. This region is currently under 
investigation by the LHC collaborations. 

Limits from  stellar cooling and primordial nucleosynthesis~\cite{Dobrescu:2004wz} are weaker than those we include in our analysis. Limits from long-range (dipole type) forces between macroscopical objects are even weaker.



Stronger constraints come from flavor physics. We include those from meson mass mixing which are the  most stringent for the processes under consideration.

\subsection{The importance of soft dark photon corrections}

Corrections due to soft dark photon exchange and emission can be important in processes with dark fermions. The strength of the coupling $\alpha_{\D}$, which we take larger than in QED, makes them sizable in the process we are interested in. 

As in QED, the decay width $\di \Gamma ^0 (s_{ij})$ for a generic $N$-body decay  is modified  by a universal  factor~\cite{Weinberg:1965nx}  that takes into account corrections from soft photons emission and we have (we follow the notation of \cite{Isidori:2007zt})
\be
\di \Gamma(s_{ij}, E)  = \Omega (s_{ij}, E)\,  \di \Gamma^0  (s_{ij}) \label{corr} \, ,
\ee
where the kinematical variables are
\be
s_{ij} = \begin{cases}
   (p_i +p_j)^2 &  i \neq 0, \, j \neq 0\\
    (p_0 - p_j)^2     &  i = 0, \, j \neq 0
\end{cases} 
\ee
with $p_i$ the momenta of the final states and $p_0$ that of the decaying particle. The corresponding  variables
\be
\beta_{ij} =\sqrt{1 - \frac{ 4 m_i^2 m_j^2}{(s_{ij} - m_i^2 - m_j^2)^2}}
\ee
can also be  defined. The energy $E$ is the maximum  energy that goes undetected in the process because of the physical limitations of the detector.

Since we are interested in factors that can compensate possible phase-space suppression in the decay, we retain only those soft-photon corrections
that  become important when the final states are produced near  threshold (in the regime where $\beta_{ij}\rightarrow 0$) and  write  \eq{corr} as
\be
\Omega (s_{ij}, E) = \Omega_C (\beta_{ij} ) 
\ee
where
\be 
\Omega_C (\beta_{ij} ) = \prod_{0<i<j} \frac{2 \pi \alpha_{\D} q_i q_j}{\beta_{ij}} \frac{1}{\exp \left[{\frac{2 \pi \alpha_{\D} q_i q_j}{\beta_{ij}}}\right] -1} \label{C}
\ee
 is the (resummed) correction due to the (dark) Coulomb interaction~\cite{sommerfeld} between  pairs of fermions with charges $q_i$ and $q_j$. 

 We  neglect all other ($E$ and non $E$-depending) soft-photon corrections because they are subleading and important only in the limit $\beta_{ij}\rightarrow 1$.


\section{The decay of neutral mesons}
\label{sec:mesons}

All neutral mesons can decay into the dark sector by means of the terms in the Lagrangian in \eq{L}. 
As promising as they would seem, the neutral pion and the $\rho$  have too short a lifetime to give a measurable BR for their decay into the dark sector.
The best candidates are  to be found in the kaon and $B$-meson sectors---even after taking into account the constraint originating in their mass mixing.  The
 $D^0$ and the charmonium  states are also  candidates but with a lower BR.

\subsection{The decay width}

 \begin{figure}[t]
\begin{center}
\includegraphics[width=2in]{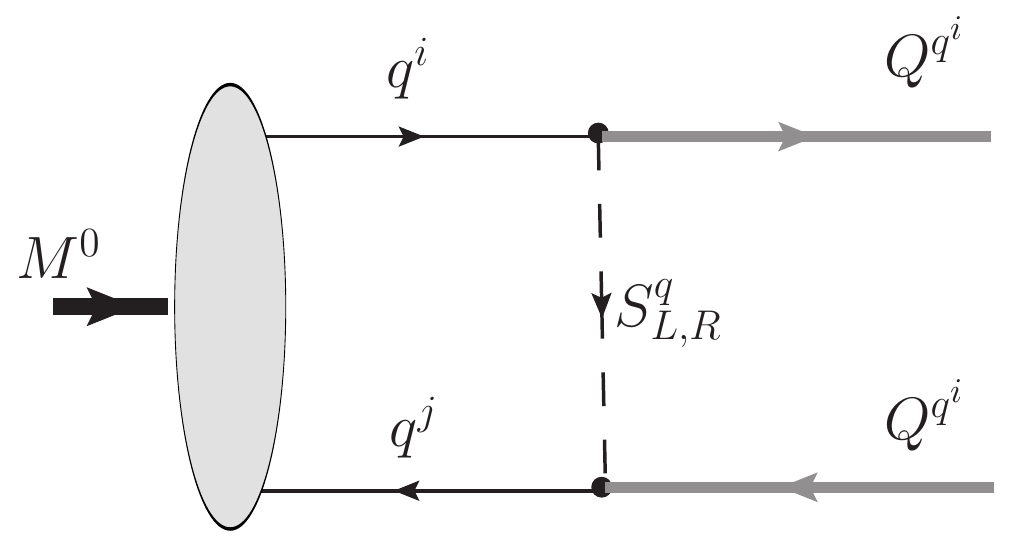}
\caption{\small The decay of a neutral meson $M^0$ ($K^L$ or $B^0$ in the text) into two dark fermions. There are two diagrams corresponding to the exchange of the two messengers $S^{Q^q}_{L}$ and $S^{Q^q}_{R}$.
\label{mesons} }
\end{center}
\end{figure}

The decay of neutral mesons can be estimated within the model of the dark sector introduced above. From the Lagrangian in  \eq{L}, after integrating out the heavy messenger fields, we can write two effective operators that give a contribution. After a Fierz transformation to bring them in a form ready to be used, they are
\bea
\hat Q_L^{ij}  &=&  \bar Q_L^i \gamma^\mu Q_L^i  \bar q_R^j \gamma_\mu q_R^i  \nn \\
 \hat Q_R^{ij}  &=&  \bar Q_R^i \gamma^\mu Q_R^i \bar q_L^j  \gamma_\mu q_L^i  \, , \label{Q}
\eea
where the indices  of the $SU(3)$ color group are implicitly summed over~\footnote{An  analysis of meson decays  with missing energy in terms of all possible effective operators---those in \eq{Q} included---is given in  \cite{Badin:2010uh}. The fermionic models they take into consideration have  BR  significantly smaller than those we find.}.

The Wilson coefficients of the two operators in \eq{Q} at the matching  are
\be
(c_L^{\D})_{ij}= \frac{g_L^{2} P_{ij}^{L}}{2 m_{\D_L}^2} \quad \mbox{and} \quad 
(c_R^{\D})_{ij}=\frac{g_R^{2} P_{ij}^{R}}{2 m_{\D_R}^2} \, ,
\ee
where the product of matrices is denoted as $P_{ij}^{L,R}=\rho^{L,R}_{ij}\rho^{L,R}_{ii}$.

The amplitude  for the neutral meson $M^0_{ij}$ decay (with $M^0_{ij}$ a bound state of $q_i \bar{q}_j$)  into dark- and antidark-fermions
$M^0_{ij} \to Q^i \bar{Q}^i$
is  given by  (see Fig.~\ref{mesons} and the operators in \eq{Q})
\be
{\cal M}^{ij}_{M^0}  =  -\frac{i}{4} \left( \frac{g_L^{2} P_{ij}^L}{m_{\D_L}^2}  
\left[\bar u_{Q^i}  \gamma_{\R}^\mu v_{Q^i}\right]  
  - \frac{g_R^{2} P_{ij}^R}{m_{\D_R}^2}  [\bar u_{Q^i} \gamma_{\LL}^\mu v_{Q^i}]  \right) 
  \,    \langle  0 | \bar q^j \gamma_5 \gamma_\mu q^i |M^0_{ij}(p) \rangle \:  \, ,
\ee
where $\gamma^{\mu}_{\LL,\R}=\gamma^{\mu}(1\pm\gamma_5)/2$, $\bar{u}_{Q^i}$ and $v_{Q^i}$ are the Dirac spinors associated to the final fermion (antifermion) states $Q^i$ ($\bar{Q}^i$) respectively, and the hadronic matrix element is given by
\be
\langle 0 | \bar q_i \gamma_5 \gamma^\mu q_j |M^0_{ij}(p) \rangle =  i f_{M^0} p^\mu \, ,
\ee
with $p_{\mu}$ the meson 4-momentum. The parameter $f_{M^0}$ for the particular meson, can be obtained from lattice estimates.

The corresponding width is  computed as 
\be
\Gamma ( M^0_{ij} \rightarrow Q^i \bar Q^i) = \frac{1}{8 \pi} \frac{|\bar {\cal M}^{ij}_{M^0} |^2|\vec k_i| }{m_{M^0}^2}  \Omega_C ( \beta_{ij}) \label{Gamma}
\ee
with
\bea
|{\cal M}^{ij}_{M^0}|^2  &=&  \frac{f_{M^0}^2 m_{M^0}^2 m_{Q^i}^2}{8 m_S^4} P_{ij}^2 (g_L^{2} +g_R^{2})^2
  \label{m2}
\eea
where $m_{M^0}$ and   $m_{Q_i}$ are the meson and dark fermion masses respectively, $q_i$ the charge of the dark fermion $Q^i$,   and  $ |\vec k_i| = m_{M^0} v_i/2$, with $v_i=\sqrt{1-4m^2_{Q^i}/m^2_{M^0}}$ the relative velocity between the dark fermions. 
The function $\Omega_C( \beta_{ij})$ is defined in \eq{C} with, in this case of two-body decay, $\beta_{ij} = v_i$.

In \eq{m2},  we have made the simplification of taking universal messenger masses $m_{\D_R}=m_{\D_L}=m_{S}$  and  $P_{ij}^{L}=P_{ij}^{R}\equiv P_{ij}$, with furthermore $\rho_{ij}^{L,R}=\rho_{ji}^{L,R}$. In the model of
\cite{Gabrielli:2016vbb} the diagonal $\rho_{ii}^{L,R}$ couplings are of order one, while the off-diagonal ones should be  $\rho_{ij}^{L,R} \ll 1$ in order to preserve the hierarchy of the CKM matrix.

\subsection{Constraints from the meson mass difference}

 A direct, and the strongest, constraint on the parameters of the model arises because the same amplitude driving   the meson decay also enters the box diagram  that gives  rise to the mass difference of the neutral meson. This quantity is given by  
  \be
 \Delta m_{M^0} = \left[  \frac{g_L^{4} (\rho_{ij}^L)^2 \rho_{ii}^L \rho_{jj}^L +  g_R^{4}(\rho_{ij}^R)^2 \rho_{ii}^R \rho_{jj}^R }{m_S^2} \right] \frac{ f_{M^0}^2 m_{M^0}}{192 \pi^2} \, , \label{DeltaM}
 \ee
 where we have used  the leading vacuum insertion approximation ($B_{M^0}=1$) to estimate the matrix element
 \be
 \langle M^0 | (\bar{q}^i_L\gamma^\mu q^j_L) \,  (\bar{q}^i_L\gamma_\mu q^j_L) |\bar M^0 \rangle = \frac{1}{3} m_{M^0} f_{M^0}^2 B_{M^0} \eta_{\mbox{\tiny QCD}}
 \ee
  and a  similar one for right-handed fields. Since we are just after an order of magnitude estimate, in \eq{DeltaM} we neglect the running (and contributions from mixing) of the Wilson coefficient $\eta_{\mbox{\tiny QCD}}$ of the 4-fermion operator. Given the long-distance uncertainties, to satisfy the experimental bound on the mass difference, we  only impose that the new contribution does not exceed the measured value (and show what happens if this bound is made more stringent).

\subsection{Branching rates for $K_L$ and $B^0$}

 The general formulas in  Eqs.~(\ref{Gamma}) and (\ref{DeltaM})  can be applied to the specific cases of interest: the invisible decays of the $K_L$ and $B^0$. 

\begin{figure}[t]
 \centering
\includegraphics[scale=0.6]{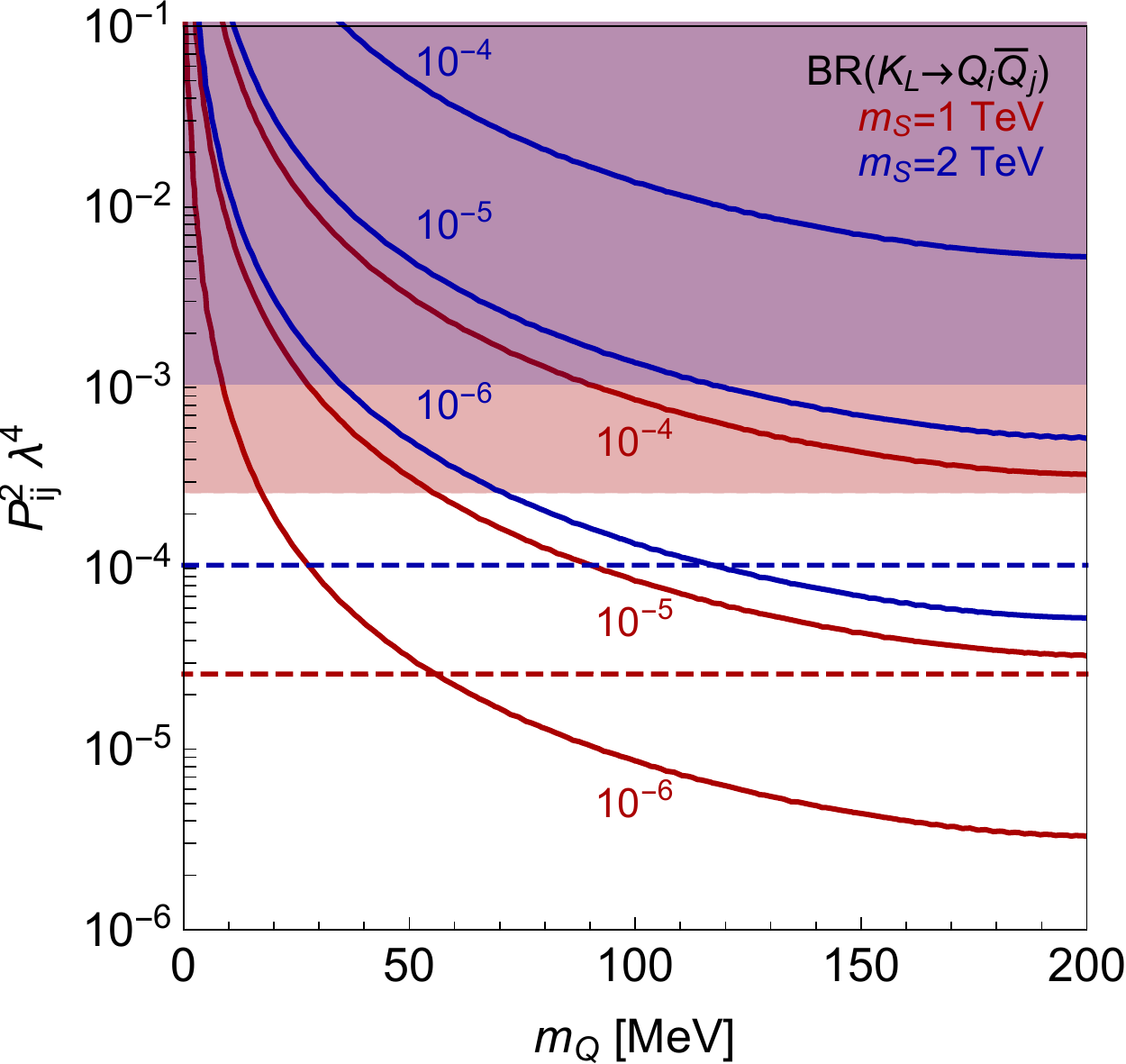} \hspace{0.7cm}
\includegraphics[scale=0.6]{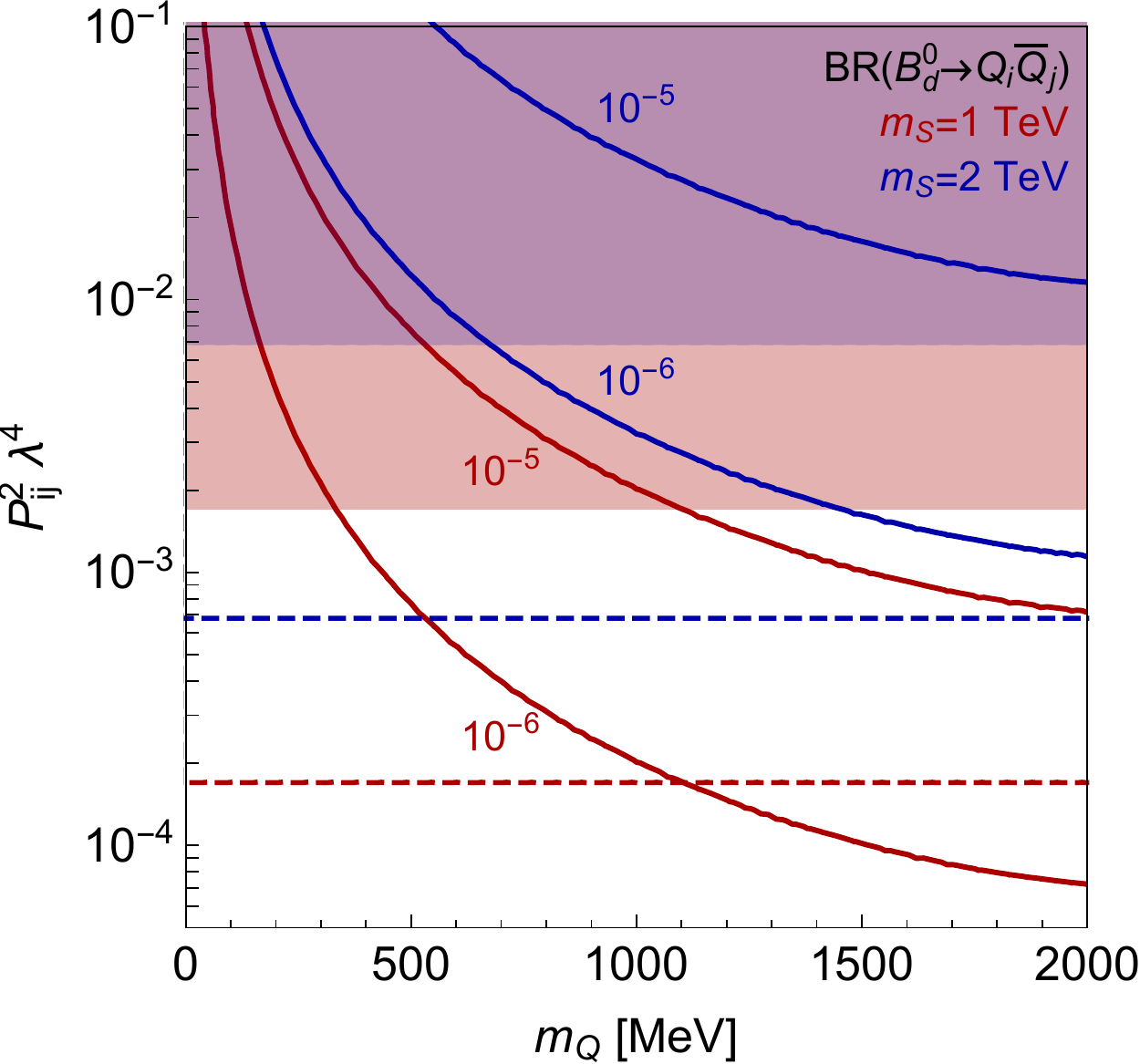}
 \caption{\small Values of the BR for the invisible decay of $K_L$ (left) and $B^0$ (right). The coupling $\alpha_{\D}$ is taken to be 0.1. Two possible choices for $m_S$ are shown. The horizontal colored bands indicate the constraint from the mass mixing for the two values of $m_S$ (red $m_S=$ 1 GeV, blue $m_S=$ 2 GeV). The case of  the same bound stronger at 10\% of the experimental limit is shown by the dashed horizontal lines. Because of chirality suppression, the width for the process goes to zero for vanishing masses of the final fermions. In the opposite limit, as the sum of these masses goes to  the threshold, the Coulomb corrections become important and keep the width finite.}
\label{plotK}
\end{figure}

    For the $K_L$ case, we have $f_{K^0}=159.8$ MeV and $m_K=497.6$ MeV~\cite{Patrignani:2016xqp}. We choose the final states to be both $Q^s$ and consider the symmetric case $g_L^i=g_R^i=\lambda$.  We take $\alpha_{\D}=0.1$ and charges $q_i=1$ to compute the  function $\Omega_C$. The total width is  $\Gamma_{K_L} = 1.287 \times 10^{-14}$ MeV~\cite{Patrignani:2016xqp}. 
    
 This BR is constrained by the mixing parameter $\Delta m_K =3.48 \times 10^{-12}$ MeV~\cite{Patrignani:2016xqp} because  the same structure enters, see \eq{DeltaM}.
Thus,  assuming that the new contribution does not exceed the experimental value $\Delta m_K$, we obtain from \eq{DeltaM}, the numerical bound 
\be
\frac{\lambda^4P_{sd}^2}{(m_S [\mbox{TeV}])^2}< 2.6 \times 10^{-4} \, . \label{dMK} 
\ee

The left panel of Fig.~\ref{plotK} shows the BR$(K_L\rightarrow Q^s \bar Q^s)$ for $\alpha_{\D}=0.1$ and a range of the parameters $m_Q$ and $m_S$. The limit from the constraint in \eq{dMK} is shown in the same plot as colored  bands. One can tighten this limit by the desired factor by rescaling the bound by the same factor: as an example, the case of 10\% of the experimental limit is shown by the dashed horizontal lines. Depending on the messenger mass, values between $10^{-4}$ and $10^{-5}$ can be reached.

 There is yet no  direct limit on this BR. An indirect value can be obtained from   the sum of all the BR of the visible decays.
The uncertainty in this sum  gives a limit of the order of $10^{-4}$. 
An experimental set-up  to bring this limit down  to  $10^{-6}$ has been proposed at the NA64 experiment at CERN~\cite{Gninenko:2014sxa}.

For the $B^0$ meson case we take the $B_d^0$  with a width $\Gamma_{B^0} = 4.33 \times 10^{-10}$ MeV~\cite{Patrignani:2016xqp}. From the lattice
  $f_{B_d}=186$ MeV~\cite{Dowdall:2013tga} while $m_{B _d} = 5279.61$ MeV~\cite{Patrignani:2016xqp}. 

As before this BR is constrained by $\Delta m_{B^0} =3.35 \times 10^{-10}\;$MeV~\cite{Patrignani:2016xqp}  thus giving 
\be
\frac{\lambda^4P_{bd}^2}{(m_S [\mbox{TeV}])^2}< 1.7 \times 10^{-3} \label{dMB}
\ee
by means of  \eq{DeltaM}.

The right panel of Fig.~\ref{plotK} shows the BR$(B^0\rightarrow Q^b \bar Q^b)$ for $\alpha_{\D}=0.1$ and a range of the parameters $m_Q$ and $m_S$. As before, the limit from the constraint in \eq{dMB} is shown in the same plot as two colored bands (and one can tighten this limit by the desired factor by rescaling the bound by the same factor: the case of 10\% of the experimental limit is shown by the dashed horizontal lines). Depending on the messenger mass,   values between $10^{-5}$ and $10^{-6}$ can be reached.

There have been several attempts to measure the invisible decay of $B^0$, both from Belle and the BaBar collaborations. The current limit is  $10^{-5}$~\cite{Hsu:2012uh}.

Our estimate indicates that, inserting values for $m_S$ still allowed by  collider searches and taking into account the constraint from flavor physics, the two BR above  fall within the explorable range of current or proposed experiments. Both decays have a SM background which is quite negligible being, as it is,  proportional to the neutrino masses squared. 
They are, literally, an open window into the dark sector that should be vigorously pursued.

\section{The decay of the neutron} 

 \begin{figure}[ht!]
\begin{center}
\includegraphics[width=2in]{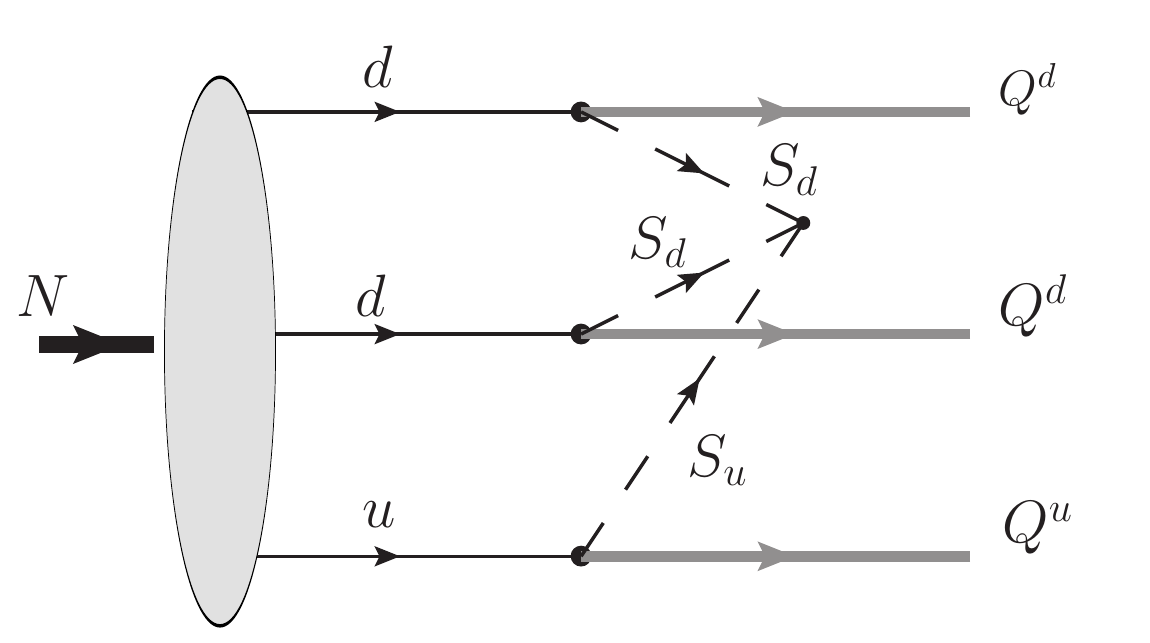}
\caption{\small The decay of the neutron $n$  into three dark fermions. There are two contributions corresponding to the two possible vertices 
in \eq{interaction} of the three scalars.
\label{neutron} }
\end{center}
\end{figure}

After integrating out the heavy messenger fields, the terms in the Lagrangians in \eq{L} and \eq{interaction} give rise to
 two effective operators violating baryon number and contributing to the decay of the neutron. They correspond to the two possible vertices among the three scalar messengers in the diagram depicted in Fig.~\ref{neutron}. These are 
\bea
\hat Q_1&=& \varepsilon_{\alpha \beta\gamma}  (\bar{Q}^{\U}_R u^{\alpha}_L) (\bar{Q}^{\D}_R d^{\beta}_L) (\bar{Q}^{\D}_L d^{\gamma}_R) \nn \\
\hat Q_2&=&  \varepsilon_{\alpha \beta\gamma} (\bar{Q}^{\U}_L u^{\alpha}_R) (\bar{Q}^{\D}_L d^{\beta}_R) (\bar{Q}^{\D}_L d^{\gamma}_R) \label{Q12} 
\eea
with $u$ and $d$ the SM up- and down-quark fields respectively. The Greek indices stand for the $SU(3)$ color group. The Wilson coefficients of the two operators at the matching  are
\be
c_1=  \frac{2 \eta_L g_L^2 g_R \rho_{\uu}^L (\rho^R_{\dd})^2 v_{\R}}{m_{\DL}^2 m_{\DR}^2 m_{\UL}^2} \quad \mbox{and} 
\quad c_2=  \frac{2 \eta_R g_R^3  \rho_{\uu}^L ( \rho_{\dd}^L)^2 v_{\R}}{m_{\DR}^4 m_{\UR}^2} \, .
\ee
where $m_{\UL}$ ($m_{\UR}$) and $m_{\DL}$ ($m_{\DR}$) are the corresponding up and down messenger masses, for the $S_L$ and $S_R$ messenger respectively.
The operators \eq{Q12} are of dimension 9 and therefore very suppressed.

For the sake of simplicity we work in a symmetric limit with $\lambda = g_L=g_R$, $\eta_L=\eta_R$ and all the messenger masses equal to $m_S$. In the same limit, the mixing matrices give a common factor  $ \rho_{uu}^L (\rho^R_{dd})^2= \rho_{uu}^L ( \rho_{dd}^L)^2 = \rho^3$ and there is a unique scale $\mu$ equal to $ \eta_L v_{\R}=\eta_R v_{\R}$. In this case, the amplitude for the decay of the neutron 
$N$
\be
N(p)\to Q^{\U}(k) Q^{\D}(k_1) Q^{\D}(k_2) 
\label{Ndecay}
\ee
is given by
 \be
     {\cal M}_N = i \frac{\lambda^3\rho^3\mu}{m_S^6} \Big( \alpha \,
[\bar{u}_{{Q}^{\D}}(k_1) \hat{P}_{L} u_{\N}(p)]  [\bar{u}_{Q^{\D}}(k_2)\hat{P}_{L} u^c_{Q^{\U}}(k)]  
+ \beta \,
[\bar{u}_{{Q}^{\D}}(k_1) \hat{P}_{R} u_{\N}(p)]  [\bar{u}_{Q^{\D}}(k_2)\hat{P}_{R} u^c_{Q^{\U}}(k)]  \Big)
\label{MNeutron}
\ee
where $p,k,k_1,k_2$ are the corresponding momenta in the reaction (\ref{Ndecay}), the chiral projectors $\hat{P}_{R/L}=(1\pm \gamma_5)/2$ and $u^c_{Q^{\U}}$ is the corresponding conjugate spinor. In deriving the above amplitude, we used the hadronic matrix elements between the vacuum and the neutron field, written as
\be
\langle 0 |   \varepsilon_{\alpha \beta\gamma} \bar{u}^{c \alpha}_R  d^\beta_L d^\gamma_R | N \rangle = \alpha \,  \hat{P}_L u_{\N}
\quad \mbox{and} \quad 
   \langle 0 |  \varepsilon_{\alpha \beta\gamma}  \bar{ u}^{c \alpha}_L d^\beta_R d^\gamma_R | N \rangle = \beta\, \hat{P}_R  u_{\N}
  \ee
 in terms of the neutron wave function $u_N$.
The coefficients $\beta$ and $\alpha$ have been estimated on the lattice to be of opposite sign and both about $0.0144 \: \mbox{GeV}^3$~\cite{Aoki:2017puj}.

The squared amplitude summed over spins and mediated over initial ones is given by  
\begin{widetext}
\be
\frac{1}{2} |\bar {\cal M}_N|^2 = 2 \lambda^6 \rho^6
\left(\frac{\eta \mu}{m_S^{6}}\right)^2\Big\{
(\alpha^2+\beta^2)(k_1\cdot k_2)(p\cdot k)\
  -2\alpha\beta m_{\N}m^2_{Q^{\D}}m_{Q^{\U}}\Big\} \, \Omega_C (\beta_{ij}) \, ,
\ee
\end{widetext}
where $m_{\N}$,$m_{Q^{\D}}$,$m_{Q^{\U}}$, are the masses of neutron, dark-fermion $Q^{\D}$ and dark-fermion $Q^{\U}$ respectively. 

The function $\Omega_C( \beta_{ij})$ is defined in \eq{C} with $q^{\U}=-2q^{\D}$  for $q^{\U}$ normalized to one; in the case of the three-body decay of the neutron,  we have
\be
\Omega_C(\beta_{ij}) = \Omega_C(\beta_{12})\Omega_C(\beta_{13})\Omega_C(\beta_{23}) \, .
\ee
This Coulomb correction requires the somewhat cumbersome definition of various coefficients. They are 
\bea
\beta_{1j}&=&\sqrt{1-\frac{4 m_{Q^{\U}}^2 m_{Q^{\D}}^2 }{(s_{1j}-m_{Q^{\U}}^2-m_{Q^{\D}}^2)^2}} \quad (j=2,3) \nn
\\
\beta_{23}&=&\sqrt{1-\frac{4 m_{Q^{\D}}^4 }{(s_{23}-2m_{Q^{\D}}^2)^2}} 
\eea
with
\bea
s_{12}&=&m_{Q^{\U}}^2+m_{Q^{\D}}^2 + 2 \,E E_2 \left(1+\beta \beta_2 \cos{\theta} \right) \nn 
\\
s_{13}&=&m_{Q^{\U}}^2+m_{Q^{\D}}^2 + 2\, E E_2 \left(1-\beta \beta_2 \cos{\theta} \right)  \label{sij}
\eea
and $s_{23}=s$. In \eq{sij} the energies are defined as
\be
E =\frac{m_N^2-s-m_{Q^{\U}}^2}{2\sqrt{s}}\,m\, , \quad
E_2=\frac{\sqrt{s}}{2}
\ee
and
\be
\beta =\sqrt{1-\frac{m_{Q^{\U}}^2}{E^2}}\, ,~~~~~\, \beta_2\,=\,\sqrt{1-\frac{4 m_{Q^{\D}}^2}{s}}\, .
\ee

The phase-space integral can be computed   in the center of mass of the two $Q^{\D}$ dark fermions. The width  is given by
\begin{widetext}
\be
\Gamma_{N \rightarrow Q^{\U} Q^{\D} Q^{\D}  } = \frac{1}{2^9 \pi^4 m_N^2} \int_{4 m^2_{Q^{\D}}}^{(m_N - m_{Q^{\U}})^2}  \di s \: \sqrt{1 - \frac{4 m_{Q^{\D}}^2}{s}} \sqrt{\left( \frac{m_N^2 - m_{Q^{\U}}^2 + s}{2 m_N} \right)^2 - s}
\int \di \Omega_\theta \left[ \frac{1}{2} |\bar {\cal M}_N|^2 \right]
  \,,
\label{int}
\ee
\end{widetext}
where  $s=(k_1+k_2)^2$ and  $\theta$ is the angle  (in this system) between $\vec k_1$ (or $\vec k_2$) and $\vec k$.
The integral in \eq{int} can be evaluated numerically.

The possibility of having  the neutron decay into the dark sector  depends on the kinematically available decay channels. If the sum of the masses of the dark fermions is smaller than the neutron mass, the decay can proceed and we can compare its rate to searches for the invisible decay of the neutron. We discuss this process in section \ref{sec:lifetime}.\footnote{See \cite{Davoudiasl:2014gfa} for an analysis of  the  decay of the neutron in states of a dark sector  plus ordinary matter.}

Since all limits on the neutron lifetime are based on  neutrons bounded in nuclei, this decay can be prevented by
choosing the dark fermion masses so as to  keep kinematically closed the decay of $^9$Be  into its unstable isotope $^8$Be. This transition has the largest energy difference (937.900 MeV) among the atomic elements and therefore closing it also closes all the other possible decays of stable isotopes. 
  
If the sum of the masses of the dark fermions just happens to be lager than 937.900 MeV but less than the neutron mass, namely 939.565 MeV,  the decay of a \textit{free} neutron remains open. We discuss this admittedly  rather artificial case   in section \ref{sec:puzzle} because of the long-standing discrepancy in the free neutron lifetime determination.

\subsection{Invisible decay of the neutron}
\label{sec:lifetime}

The absence of an invisible decay of neutrons in $^{16}$O and $^{12}$C  from SNO~ \cite{Ahmed:2003sy} and KamLaAND~\cite{Araki:2005jt}   put a  stringent limit of  $\tau_N^{\tiny \mbox{inv}} \gtap 10^{29}$ years on such a channel. 

The operators in \eq{Q12} are dimensionally suppressed and therefore naturally provide a width that can be very small. 
For instance, for dark fermion masses 
$m_{Q_{\U}}=m_{Q_{\D}}=100$ MeV, by means of \eq{int}  we find that
\be
\Gamma_{N \rightarrow \mbox{\tiny invisible}}  \simeq  4.9 \times  10^{-55}  \left( \frac{\lambda \rho}{4 \pi} \right)^6 \left( \frac{\mbox{100 TeV}}{m_S} \right)^{10}\mbox{GeV} \label{lifetime}
\ee
for  $\eta \mu = 0.1 \,m_S$ and $\Omega_C=1$ (because we are far from the production threshold). The width in \eq{lifetime} must be smaller than $10^{-61}$ GeV to satisfy the lifetime bound---which is achieved  for couplings $\lambda\rho \sim 1$ and $m_S \sim 100$ TeV. 

Different values for the masses of different messengers make the estimates  in \eq{lifetime} for the neutron decay and those for the meson decay in section \ref{sec:mesons}  compatible.

The operators in \eq{Q12}  provide an interesting example of  operators violating the baryonic number that can live at a scale of order 100 TeV---and therefore much smaller than the typical GUT scale---without further assumptions on the size of the dimensionless couplings. 
The result in \eq{lifetime} shows that the study of the neutron invisible decay provides a promising test for the disappearance of ordinary matter  into the dark sector.

\subsection{The neutron lifetime puzzle}
\label{sec:puzzle}

The lifetime of the neutron has been measured by counting  either cooled neutrons stored in a container (the bottle method)~\cite{Patrignani:2016xqp}, see \cite{Serebrov:2017bzo}  for the most recent determination, or protons coming from neutron decaying while traveling  in a given volume (the beam method)~\cite{Byrne:1996zz,Yue:2013qrc}.
The two measurements do not agree and the discrepancy (the beam result is about 8 seconds longer) has a significance of nearly 4$\sigma$.

A very interesting  explanation would be the existence of  an additional invisible decay channel of the neutron, 
as  proposed in~\cite{Fornal:2018eol}, which will affect the beam method measurement but not the bottle method. In particular, the authors of~\cite{Fornal:2018eol} assumed a dark decay of the neutron either  into a dark fermion and a photon
 or into a dark scalar  and a dark fermion. This possibility was further elaborated in~\cite{Cline:2018ami}.\footnote{See, also,  \cite{Leontaris:2018blt,Berezhiani:2018eds} for neutron decay in the context of  neutron-antineutron oscillations.} The lifetime of the neutron is related to the axial coupling determination~\cite{Czarnecki:2018okw}.

Given the Lagrangian in \eq{L} and \eq{interaction}, the decay of the neutron into the dark sector, within the  model  we have introduced, can only take place by means of the diagram in Fig.~\ref{neutron} with the neutron decaying into three dark fermions. This is not one of the processes previously envisaged either in ~\cite{Fornal:2018eol} or \cite{Cline:2018ami}. The charge conservation built in the model prevents a similar decay for the proton.

Astrophysical bounds from the dynamics of neutron stars~\cite{McKeen:2018xwc} do not rule out this possibility because of \textit{Pauli blocking}.  This is the same mechanism that prevents  neutrons in a neutron star  to $\beta$-decay.
In a neutron star all the fermions are mostly in a degenerate state. After the neutron decay has started, the presence of $N$ dark fermions  gives rise to the corresponding Fermi energy 
\be
E_F= \frac{1}{2 m_Q}  \left( \frac{3 \pi^2 N}{V} \right)^{2/3} \simeq 10^{-19}  \frac{N^{2/3}}{m_Q} \, \mbox{MeV}^2 \, ,
\ee
where $V$ is the volume of the neutron star, which we take to have a radius of about $10^4$ meters.
When $E_F$ is  larger than   the energy available in the decay (about 1 MeV),  further neutron decays are  effectively stopped. This happens after (for $m_Q\simeq 100$ MeV) about $10^{52}$ decays, that is after 1 out of $10^{5}$ of the neutrons in the star have decayed. This is too small a fraction to appreciably change the equation of state of the neutron star,  change its mass limit and activate the bounds in \cite{McKeen:2018xwc}.

The double limit imposed by the $^9$Be stability and the mass of the proton 
\be
937.900 \; \mbox{MeV} < 2 m_{\D}+ m_{\U}< 939.565 \; \mbox{ MeV}
\ee
makes for a very narrow window  where the sum of the masses of the dark fermions must be. 

In this region the limits from the neutron lifetime discussed in the previous section need not apply  (the decay is closed by the beryllium bound) and $m_S$ is only constrained by the LHC data. 

The  nearness of the sum of these masses to the neutron mass gives a  very strong suppression in the  phase space of the decay (of about 4 orders of magnitude), only partially  compensated by the  enhancement due to the Coulomb  interaction of the final states (which is partially suppressed by the repulsive component and about 1 order of magnitude).

For $\alpha_{\D}=0.1$, $m_{Q^{\U}} = m_{Q^{\D}} \simeq 313$ MeV (to satisfy the nuclear physics constraints),   and after taking $\eta \mu = 10\, m_S$---at the very limit of the unitarity constraint---we find 
\be
\Gamma_{N \rightarrow Q^{\U} Q^{\D} Q^{\D} } \simeq 4.9 \times 10^{-35}  \left( \frac{\lambda \rho}{4 \pi} \right)^6  \left( \frac{1 \; \mbox{TeV}}{m_S} \right)^{10} \: \mbox{GeV} \, . \label{nnn}
\ee

For the width in~\eq{nnn} to be of order $10^{-30}$ GeV---the value necessary to explain  the  discrepancy in the neutron lifetime data---we must take 
$m_S$  around 200 GeV, a value still  allowed by the LHC data if the messenger decays almost exclusively into a top quark, and $\lambda \rho \simeq 6$. 
 This is the extreme choice for the model parameters alluded in the abstract. If (most likely, when) the LHC will close this window, the neutron lifetime puzzle will no longer be explained by the model of the dark sector we consider here.

\begin{acknowledgments}
We thank Jessie Shelton for bringing to our attention the possibility in our model of  Pauli blocking the neutron decay in neutron stars.
\end{acknowledgments}



\end{document}